\documentclass[12pt]{article}
\input{epsf.tex}

\begin{document}

\begin{center}

{\bf POSSIBILITY OF EXPERIMENTAL DETERMINATION OF RELIABLE PARAMETERS
OF THE COMPOUND-STATE GAMMA-DECAY AND SOME ERRORS OF ANALYSIS:
$^{96}$Mo AS AN EXAMPLE}\\\end{center}
\begin{center}
{\bf A.M. Sukhovoj, V.A. Khitrov}\\
\end{center}\begin{center}
{\it  Joint Institute
for Nuclear Research, 141980, Dubna, Russia}\\
\end{center}

Comparison between potential possibilities and inevitable systematic errors
of one- and two-step reactions for obtaining of maximum reliable data on
level density and radiative strength functions after decay of excited
levels of complicated nuclei has been performed. It was shown that the use
for this aim of two-step reactions instead of one-step reactions provides
for potential possibility to decrease systematical errors of mentioned
nuclear parameters, as minimum, by several times.

\section{Introduction}\hspace*{16pt}

Level density  $\rho$ and radiative strength functions 
$k=f/ A^{2/3}=\Gamma/(E_\gamma^3 D_\lambda A^{2/3})$ of cascade gamma-transitions
following decay of high-lying levels ($E_{ex}>5-10$ MeV) for main portion
of excited nuclei can be determined experimentally only from solution of
reverse task -- from the spectra of cascade gamma-decay measured with
``bad resolution". There is standard task for mathematical analysis.
Usually, it has infinite number of possible solutions.
But, the interval of the possible $\rho$ and $k$ values can be both infinite
or strongly limited by experimental conditions.
In the first case, determination of $\rho$ or $k$ is impossible without
applying additional information, in the second -- it is possible to determine
the small enough region of the $\rho$ and $k$ values which reproduce
experimental spectra with the value $\chi^2/f <1$.
In the second variant it is appeared, as minimum, possibility
to exclude from consideration some nuclear models which do not provide
required precision in reproduction of experimental data.

Such models of $\rho$ and $k$ inevitably appeared by theoretical analysis
of experimental data obtained earlier with large systematical errors.
The latter unambiguously follows from comparison of $\rho$ and $k$ values
determined in experiments of different types. The primary problem in this
comparison and following selection of nuclear models is realistic
estimation of systematical uncertainties of the data under consideration.

\section{Experiments of different type}\hspace*{16pt}

Qualitative difference in potential possibilities to get reliable data on
$\rho$ and $k$ is caused, first of all, by the type of experiment.
The usual (below -- one-step) experiment corresponds to registration of
given reaction product and, main, to comparison between probability
distribution of its emission and tested reaction notions independently
on probable yield of other products.

Two-step experiment assumes both measurement and comparison of the measured
and calculated
with the tested functions $\rho=\psi(E_{ex})$ and $k=\phi(E_\gamma)$
emission probability of two correlated reaction products.
There can be two gamma-quanta or particle and gamma-quantum.
The function shape for probability distribution of their
registration with the determined parameters in one- and two-step reactions is
different. Just this difference causes different value of systematical
uncertainties of measured nuclear parameters and high resulting reliability
of their values in two-step reaction.

The reason for considerable discrepancy between achieved precision in
determination of experimemtal data on $\rho$ and $k$ is easily revealed at
comparison between one- and two-step reactions. 
In the first case, determination of these nuclear parameters is performed
usually from the spectra whose $S$ amplitudes are described by expression
like 
$S\propto \rho k/\sum (\rho k)$. Registration probability $B_r$ of the
next reaction product (here -- gamma-quanta
to given final levels) has another form of
dependence on the same parameters: $B_r \propto k/\sum (\rho k)$.
This leads to principle change in values of systematical errors
$\delta \rho$ and $\delta k$ -- decrease of their values by several
times as compared with one-step reactions.

\section{One-step reaction}\hspace*{16pt}

Authors \cite{Morea} performed re-analysis of their experimental data
(practically -- one-step reactions ($^3$He,$^3$He$^{'}\gamma$),
($^3$He, $\alpha\gamma$)) and changed small portion of results presented earlier.

On the whole, principle discrepancy between physical picture of cascade
gamma-decay obtained in Oslo and Dubna remains.
Characteristic feature of the first data set -- smooth change in level
density as changing excitation energy; sharp changes in nuclear properties
are absent. Any correlation between level density and emission probability
of gamma-quanta is not reported.

In the second case is observed alternative picture of processes occurring in
nucleus.

Approximation of level density \cite{PEPAN-2006} by Strutinsky model
\cite{Strut} and sum of radiative strength functions by semi-phenomenological
dependence \cite{Appr-k} gives the following picture of gamma-decay:

(a) level density below neutron binding energy $B_n$ is determined by quantity
of excited quasi-particles breaking threshold of 3-5 nucleon pairs.
This number depends on shape of energy dependence of the nucleon pair
correlation function in heated nucleus. Main portion of excited levels below
$\approx 0.5B_n$ is caused by nuclear phonon excitations and (for deformed
nuclei) by their rotation. This result qualitatively coincidences with the data
on coefficients $K_{\rm coll}$ of collective enhancement of level density
presented in file  \cite{RIPL};

(b) sum of radiative strength functions of E1- and M1-transitions is
described
very well
by superposition of model \cite{KMF} and local ``peak"~structure.
The latter has maximum value in region of levels whose wave functions contain
large few-quasi-particle components or large phonon components (excitation
energy region with $K_{\rm coll} \gg 1$) with the ``tail"~  potentially
decreasing as decreasing gamma-transition energy. Moreover, functional
dependence \cite{KMF} has the weight  $w \sim 1/K_{\rm coll}$ and,
correspondingly, less nuclear temperature $T$. Parameters $w$ and $T$ depend
on parity of nucleon number in nuclei studied up to now.

Therefore, there is urgent necessity in further analysis of systematical
errors of different methods for determination of $\rho$ and $k$.
In \cite{TSC-err} is demonstrated  that change in total intensity of
two-step cascades by $\pm 25\%$ from the value obtained in experiment does
not lead to principle change in form of determined $\rho$ and $k$ values.
The use of hypothesis \cite{Axel,Brink} instead of experimental function
$k(E_\gamma,E_{\rm ex})$ overestimates the obtained  level density,
most probably, not less than by two times (the values of $k$ are
underestimated, respectively). The other ordinary systematical errors
have minor value. On the whole, difference between the pictures of
gamma-decay obtained in different experiments cannot be related mainly to
systematical errors of method \cite{PEPAN-2005}.

Necessity, direction and considerably larger volume of required re-analysis
in \cite{Morea}, as it should be expected, are caused by inaccurate
subtraction of Compton background (and other backgrounds) in each experimental
total gamma-spectrum and errors of normalization of all the spectra to the
same number of decays.

Unfortunately, authors of \cite{Morea} did not performed quantitative
analysis of systematical errors and, main, coefficients of their transfer to
the determined parameters. In particular, relative normalization of total
gamma-spectra was done with the use of experimental multiplicity of
coincidences. As it follows from the works published by Oslo group, probable
discrepancy between its value and mean number of gamma-quanta in cascade
following decay of levels with excitation energy $E_{ex}$ was not investigated.
Nevertheless, determination  of systematical errors of the obtained $\rho$ and $k$
requires one to estimate possible errors of the measured intensity and, main,
coefficients of their transfer to the ``first generation spectra". And then --
to the determined parameters $\rho$ and $k$.

\subsection{Error transfer and possibility to estimate its coefficients
for one-step reaction}\hspace*{16pt}

Qualitative notion on the considered values can be obtained in different ways:

1. Direct comparison between  ``raw"~  and ``primary"~  spectra presented in the
same scale. (Normalization of these spectra relative to their high-energy
parts can be performed unambiguously). This allows one easily to estimate
required precision in determination of intensity of total gamma-spectra
following decay of levels lying in region of neutron binding energy (and lower)
for any nucleus studied in Oslo. For the low-energy primary gamma-transitions
(for example, $E_\gamma \approx 1-2$ MeV), relative error of subtracted
Compton background of raw-spectrum must not exceed $\sim 0.01$.
(It is obtained at rather optimistic estimation of required precision in
determination of the primary gamma-transition intensities - 10\% and exceeding
of ``raw"~spectrum above the ``primary"~one only $\sim 10$ times at low energy
of gamma-quanta).

Therefore, inevitable and different components of background (independently
on their nature) and any distortion at registration of gamma-quanta by
scintillation detector must be small enough (i. e., their contribution in
low-energy part of total spectrum must be noticeably less than $\sim 1\%$).
In the other case they must be determined independently on main experiment,
at least, with the same precision.

2. Calculation of total gamma-spectra for different model functions of $\rho$, $k$
and following comparison of differences $\Delta S_{ij}^{cal}$ with corresponding
values of $\delta \rho_{ij}$ and $\delta k_{ij}$ for any possible pairs of
$\rho$ and $k$.

\begin{figure}[htbp]
\vspace{4cm}
\leavevmode
\epsfxsize=18cm

\epsfbox{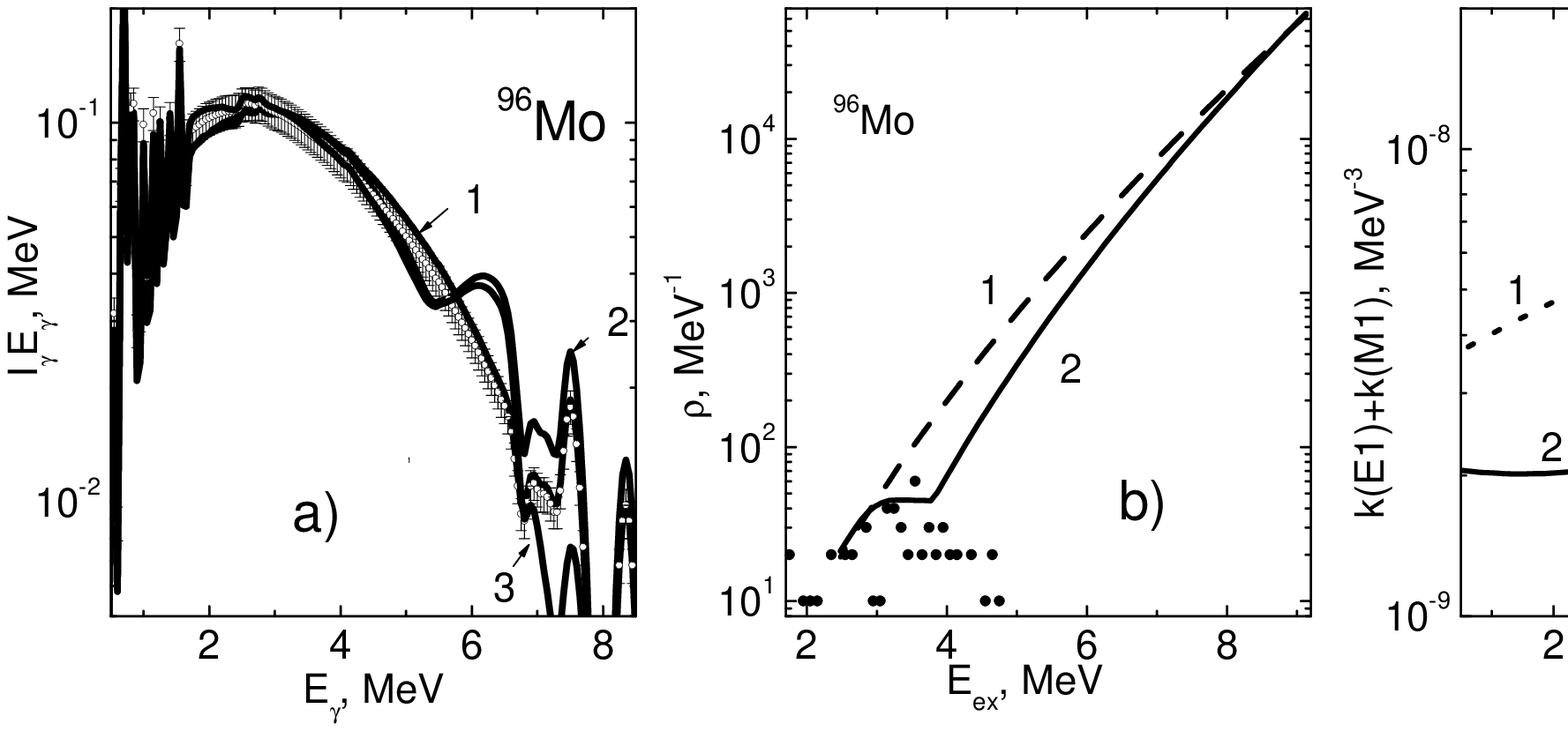}
\vspace{-11cm}

Fig. 1a). 4 variants of the calculated total gamma-spectra in $^{96}$Mo.
Points with errors -- calculation for models \cite{KMF} and \cite{BSFG},
curve 1 - \cite{Axel,Brink}+\cite{BSFG}, curve 2 -- \cite{KMF} and step-like
level density from Fig.1b), curve 3 -- \cite{Axel} and step-like structure.
Fig. 1b). Points -- total number of levels in known decay scheme
\cite{ENSDF}, curve 1 - \cite{BSFG},
curve 2 -- level density with step-like structure. Fig. 1c).
Curve  1 - data \cite{Axel}, curve 2 --\cite{KMF} together with $k(M1)=$const.
\end{figure}

In Fig. 1a) are presented four total gamma-spectra for $^{96}$Mo calculated
within two models of level density and two models of radiative strength
functions shown in Fig. 1b and Fig. 1c), respectively. For convenience of
comparison, calculated spectra are presented in form $S=I_\gamma E_\gamma$.
For one spectrum there are given errors of calculation which are equal to 10\%.
As it is shown below, the total gamma-spectrum can be presented by sum of
spectra of the irst generation gamma-transitions depopulating low-lying levels. Id est,
it follows from the fact of small difference between model calculation in
low-energy part of spectrum that the required total error in determination
of total gamma-spectra in method \cite{NIM}, most probably, must be
considerably less than 10\%.

3. Calculation of spectra $h$ of primary gamma-transitions for different
excitation energy of studied nucleus and given parameters $\rho$, $k$ with
folding of these functions in total gamma-spectra $S$ according to expression:

\begin{equation}
S_i=h_i+\sum_l(h_l S_l).
\end{equation}

The following distortion of spectra $S_i$ by different errors $d$ and
reconstruction of distorted values of $h_i$ by means of reverse to (1)
procedure:

\begin{equation}h_i=S_i d_i
-\sum_l(h_l S_l d_l)
\end{equation}
allow easy modeling of  error transfer of total gamma-spectra to the errors
of the primary gamma-transition spectrum. Necessity and possibility of
effective search for error transfer of the total gamma-spectra to the
primary transition spectra was first investigated and suggested in
\cite{Oslo-err}.

(a) Identical analysis for gamma-decay  of  $^{96}$Mo excited levels shows,
for example, that \cite{Morea} ``unexpected enhancements in the radiative
strength functions (RSF) of low energy gamma-ray..."~ is, probably,
inevitable consequence of ordinary systematical errors in normalization of the
total gamma-spectra. For instance, at linear distortion of their area by
coefficient  $d=1+d_{max}((B_n - E_{ex})/B_n)$
in energy interval of
decaying levels from 9 MeV to ground state. 
There were tested two variants of distorting function ($d_{max}=10\%$ and
$d_{max}=-10\%$). The results are presented in Fig. 2.
Most probably, this estimation
of possible systematical error $d_{max}=\pm 10\%$ is optimistic and underestimated by several
times. This conclusion was made on the ground of dispersion of experimental
multiplicity of gamma-quanta presented in \cite{PRC-1972}. Just this
parameter of gamma-decay is used for relative normalization of the total
gamma-spectra by group from Oslo.

Moreover, double overestimation of intensity of the primary transition
spectrum is observed at the primary transition energy of about 1 MeV for
the nuclear excitation energy $E_{ex}= 9$ MeV (Fig. 2).
Analogous overestimation is regularly present and at lower excitation energy
$E_{ex}$. This error quickly increases at the less than 1 MeV primary
transition energy $E_1$. At higher energy -- changes sign and magnitude.
Transfer coefficients of these errors to the determined level density and
radiative strength functions strongly exceed analogous values in analysis of
two-step cascade intensities \cite{TSC-err} due to difference of functional
dependences on $\rho$ and $k$ of experimentally measured distributions.

\begin{figure}[htbp]
\vspace{4.5cm}
\leavevmode
\epsfxsize=17cm

\epsfbox{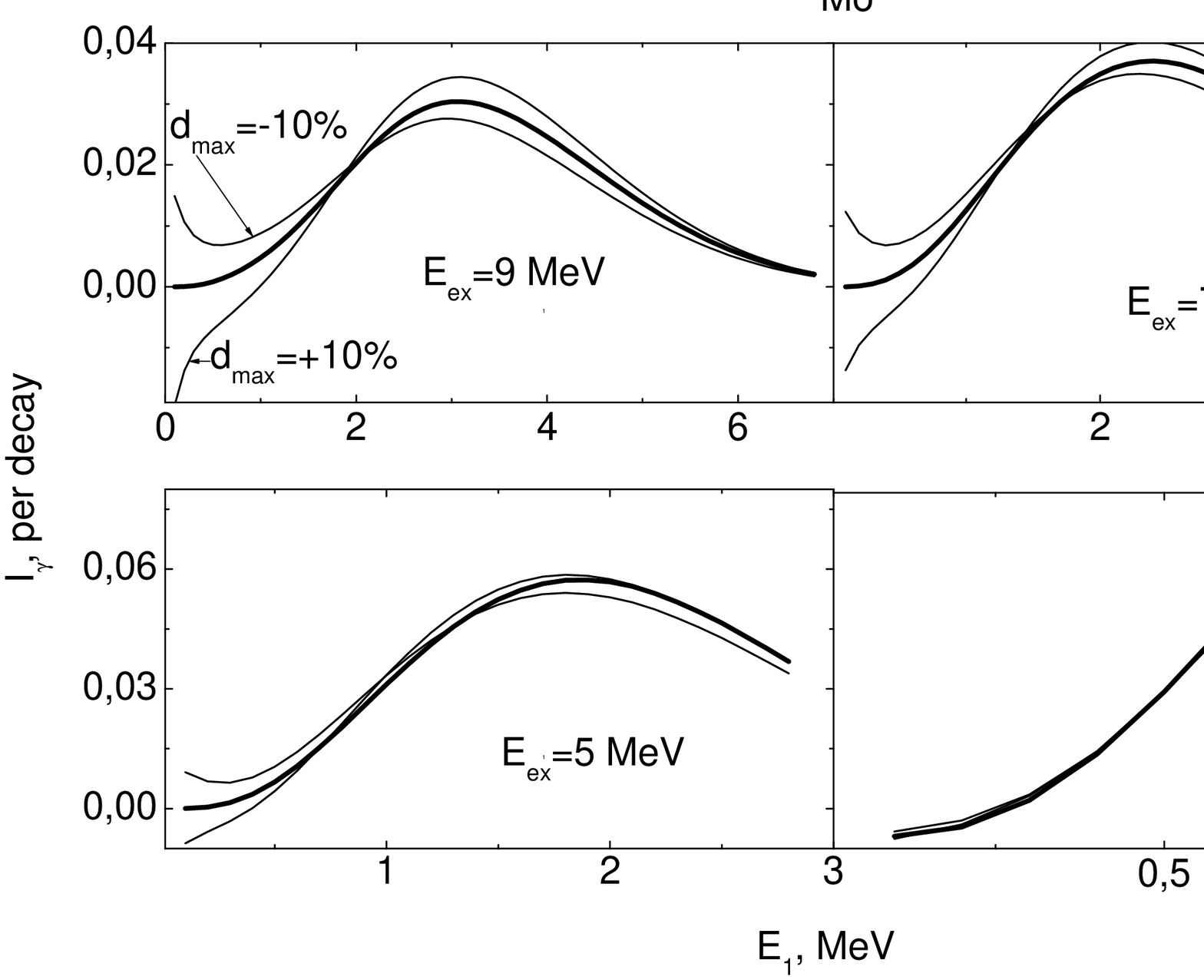}
\vspace{-7cm}

Fig. 2. Solid curve -- model calculation of the primary gamma-transition
intensity in $^{96}$Mo. Two thin curves show its change for linear
overestimation (underestimation) of area of total gamma-spectra for different
energy $E_{ex}$ of decaying levels.
\end{figure}

(b) There are no doubts in difference of shapes and areas of incoming in (2)
gamma-spectra following depopulation of levels of the same energy but excited
by primary gamma-transitions from higher-lying ($S_i$) levels or in result of
the nucleon product emission of nuclear reaction ($S_l$).

This is caused by both difference in energy dependence of the primary E1- and
M1-transitions (directly observed in method \cite{PEPAN-2005}) and widening of
spin interval excited by dipole gamma-transitions. The method for determination of
corresponding error is unknown. 

However, it should be taken into account that the values of analyzed errors
and coefficients of their transfer to values of level density and radiative
strength functions must be determined numerically for the worst cases.

(c) Multi-parametric fitting is usually performed using the method suggested
for the first time by Gauss and then developed by other mathematicians for the
case when corresponding system of equations is singular or close to singular.
The method consists in the following: the value of vector-column $X$ consisting
from $n$ parameters is determined (for example, \cite{FUM}) in the vicinity of
their
actual values for $k+1$ iteration by the matrix equation
\begin{equation}
X_{k+1}=X_{k}-(J^{T}GJ)^{-1} J^{T} G S(X_{k}),
\end{equation}
where $G$ is the matrix of weights, the Jacobi matrix $J$ and corresponding
transposed matrix  $J^{T}$ are the matrixes of derivatives from function
$S(E_\gamma)$  with respect to the desired radiative strength function $k$ of
transition and number $\rho$ of levels in a given energy interval
$\Delta E$ of corresponding spectrum. $S$ is the vector-row of $m$ experimental
points in all spectra involved in fitting of parameters. It is obvious that
eq.~(3) has a solution (unique) only upon condition of existence of covariant
matrix $C=(J^{T}GJ)^{-1}$. Otherwise, process (3) is realized using some
type of regularization. Existing programs of multi-parameter fitting can
find some arbitrary solution of system in case when system of equations is
degenerated.

In the case considered here \cite{Morea}, matrix $C$ is degenerated \cite{Gauss-err} even
at the use of all available spectroscopic information (known level density
in two excitation energy points, total radiative width in vicinity of $B_n$
and ratio $k(M1)/k(E1)$ near $B_n$).

As a consequence, iterative process for search of maximum of likelihood
function requires one to use regularization (increasing of diagonal elements of
matrix $C$), which does not distort direction along gradient of the likelihood
function.

Compulsory limitation of any elements of corrective vector $\delta X/X$ for
unknown $\rho$ and $k$ by relative value $P$ in limits $0.01\leq P\leq 0.2$ 
(expression (17) in \cite{NIM}) at each iteration deflects this vector from
the maximum of the likelihood function. Authors \cite{NIM} did not present
proof for convergence of the process under conditions listed above.
Therefore, one can conclude that the maximum of likelihood function was not
achieved.

As it follows from accumulated experience of determination of function
parameters (presented in textbooks on mathematical statistics), they should
be determined at maximal variation of initial values of level density and
radiative strength functions. Under conditions of degenerated matrix, maximum
of likelihood function cannot be the only. Conclusion completely contradicts
algorithm \cite{NIM}.

(c) High sensitivity and very large volume of accumulated information on
two-step reaction $(n,2\gamma)$ \cite{PEPAN-2005} allows one to get
principally new information on structure of nuclei of any type in considerably
wider excitation energy interval than it is available for classical
nuclear spectroscopy. Hence, reliable parameters of gamma-decay can be
obtained only under condition of accounting for the strongest violation of the
Axel-Brink hypothesis for gamma-transitions to the levels with different ratio
between vibrational and quasi-particle components. It is regularly and easily
revealed experimentally as very significant enhancement in cascade population
of levels in region of step-like structure (nuclear excitation energy -- several
MeV) in investigation of two-step cascades. It should be noted, that accounting
for probable dependence $k(E1)+k(M1)=f(E_\gamma ,E_{ex})$ simultaneously
decreases discrepancy between the calculated and experimental total
gamma-spectra also \cite{TotSpe}).

Analysis \cite{Fe57} showed principle discrepancy between shape of
experimental intensity of two-step cascades in $^{57}$Fe, for example, and
that calculated with the use of the data on the  ($^3He,^3He^{'} \gamma)$
reaction. The analysis was performed with accounting for all requirements
of mathematical analysis and mathematical statistics. 

\section{Two-step reaction}\hspace*{16pt}

The term ``two-step"~reaction assumes,
in general,
experimental measurement of product of
partial cross-sections for two successive products of nuclear reaction in case
when spectrometer resolution is enough for observation of individual peaks,
or  sum of their intensities over excitation energy region of intermediate
levels -  at bad resolution. Besides, there is possible registration of
charged particle and following gamma-quantum. Reaction $(n,\gamma\alpha)$
studied early in FLNP JINR also belongs to class of two-step reactions.

Naturally, it is necessary to account for possibility of significant
systematical error in strength function of low-energy primary gamma-transitions
determined in analysis. This error can be due to incorrectness of hypotheses
used for experimental data analysis and, correspondingly, can lead to
incorrect parameterization of model \cite{KMF} obtained on this basis.
(For example, at considerable increase in  $\alpha$-widths relative to
the average.) Most probably, there is the main problem for analysis of
two-step reactions and  source of main systematical uncertainties in nuclear
parameters obtained from them. Main sources of errors in processing of the
two-step cascade intensities were analyzed in \cite{Oslo-err} and detailed in
\cite{Fe57}. 

The used in \cite{PEPAN-2005} main principles and algorithms for the
two-step cascade intensity analysis were developed in functional analysis
and mathematical statistics and completely correspond to their basis notions.

Practically revealed sources of systematical errors and problems which
should be solved by analysis of two-step cascade intensities are the
following:

1. The assumption \cite{Axel,Brink} on absence or weak influence of nuclear
structure on its main parameters -- level density and emission probability
of gamma-quanta with the same energy but different energies of nuclear
excitations is obviously mistaken.

2. The existence of false solutions of reversed task considered here is
inevitable. 

3. There is necessary to take into account the strongest correlation between
both different desired parameters and parameters of the same type
(but for different nuclear excitation energy).

4. Non-linearity of error transfer coefficients of the measured spectra
and their change at increasing (decreasing) error of experiment
determines the width of interval for the possible $\rho$ and $k$ values
for different systematical errors of cascade intensities and and vice versa.

These statements are based on modern theoretical ideas, the set of available
experimental data and general methodological scientific principles.

1. There are:

a) Analysis of fragmentation of nuclear states of different complexity
\cite{MalSol} showed strong irregularity of this process.
At any nuclear excitation energy, wave functions of levels can contain
large components of different type. They can be in matrix elements of
gamma-transitions and penetration coefficients of nuclear surface for nucleon
products of nuclear reaction. This contradicts main postulates of ``statistical"
model of nucleus.

b) The same conclusion follows from coefficients of vibrational enhancement
of level density \cite{RIPL} in the region of neutron binding energy and
estimation of parameters of the primary gamma-transition intensity distribution
in reaction $(\overline{n},\gamma)$. The achieved level of experiment and
modern mathematics apparatus allows its treatment without the use of this
obsolete postulate.

c) Only the comparison between the parameters of nuclear reaction obtained
in this way and theoretical ideas can give objective picture of processes
occurring in nucleus.

2. Because simultaneous
extraction of $\rho$ and $k$ from untransformed experimental spectra of
two-step reactions always gives some set of false solutions.
Region of their values
must be minimal at registration of different products at the first and second
steps of reaction. But, it very strongly increases at registration of two
gamma-quanta with lifetime of intermediate level in femtosecond diapason
by any known spectrometers of gamma-coincidences.

In consequence, experimental spectra of the $(n,2\gamma)$ reaction can be
reproduced with $\chi^2/f< 1$ by infinite number of the level density
and radiative strength
functions -- on gamma-quantum energy and structure of levels connected by
gamma-transition. Moreover, ratio between the obtained maximal and minimal
values of $\rho$ and $k$ can exceed some tens \cite{Oslo-err}.

Very essential reduction of interval of their possible values
in the $(n,2\gamma)$ reaction requires one
to determine the portion of the primary transition intensity in arbitrary
energy interval of cascade gamma-transitions in vicinity of chosen energy
$E_\gamma$.
This task can be solved \cite{Prim} with acceptable error by accounting for
shape of line (changes in intensity and number of registered cascades) of
primary gamma-transition with different energy  $E_1$.

3.  Any change in function $\rho$, for example, precisely reproducing
experimental spectra results in adequate change of the same value for
 other excitation energy or/and strength functions $k$.
This correlation is realized through the total radiative widths of initial
and intermediate cascade levels.
 Such  correlation is clearly observed in method
 \cite{PEPAN-2005} for all the nuclei studied in Dubna. This follows
 from the fact that
in experiments with even ordinary detectors is
observed some tens percent of the total intensity of the primary
gamma-transitions 

\begin{equation}
 I_{\gamma\gamma}(E_1)=\sum_{\lambda ,f}\sum_{i}\frac{\Gamma_{\lambda i}}
{\Gamma_{\lambda}}\frac{\Gamma_{if}}{\Gamma_i}.
\end{equation}

The analysis performed, for example, in \cite{LA172,Bud57,Pra198} ignores
strong correlation between intensity of any cascade with other gamma-quanta.
Therefore, comparison only of central parts of the experimental spectra with
different variants of model calculation by discrepancy at the ends of spectra
guaranties absolute unreliability of the made conclusion.

4. Applicability of some set of the $\rho$ and $k$ models for reproducing
experimental cascade intensity by means of criterion $\chi^2$ was estimated
without accounting for significant nonlinear coefficients of error transfer
of experimental spectrum to errors of parameters. They significantly differ
from unit. In this situation, the width of confidence interval for errors of
tested level density and radiative strength functions can be unlimited large
at least in some cases.

The examples of the two-step cascade intensity analysis in $^{57}$Fe,
$^{172}$Yb, $^{163}$Dy
and $^{198}$Au performed without accounting for mentioned above specific of
two-step reaction $(n,2\gamma)$ can be found in \cite{LA172,Bud57,Pra198}.
Accounting for this specific \cite{Fe57,Oslo-err,Au198} gives significantly
different data on both level density and radiative strength functions and does
not contradicts our data for other nuclei.

Nucleus $^{96}$Mo is not exclusion, as well.

\subsection{Grounded and ungrounded conclusions at analysis of experiment}
\hspace*{16pt}

Necessity in estimation of ground of experimental conclusion unambiguously
follows from two examples:

a) comparison between results of different experiments at test of the
Bohr-Mottelson \cite{BM} (or Axel-Brink) hypothesis of independence of reverse
reaction cross-sections on wave function structure of excited level of final
nucleus and

b) logic in choice of conclusion on the nuclear process picture at presence
of infinite number of possible parameter values concentrated in final interval
of their possible magnitudes.

1. Extraction of level density from spectra of one-step reactions is impossible
without the use of hypotheses of independence of reverse reaction cross-section
on excitation energy of final nucleus. Corresponding error of any adopted
hypothesis is directly transformed into unknown systematical uncertainty in
determination of $\rho$.

This problem is not so important for two-step reactions. As it was shown in
analysis of change in cascade population of levels of studied nucleus below
$0.5B_n$ \cite{PEPAN-2005}, deviation of cross-section from general trend at
different nuclear excitation energy in reaction $(n,2\gamma)$ has different
sign. Qualitatively, the effect of sign-changeable cycling in deviation of
cross-sections from averaged dependence can be interpreted in frameworks of
theoretical conclusions about regularities of fragmentation process of states
with different numbers of quasi-particles and phonons. So, inapplicability
of the Axel-Brink hypothesis for gamma-quanta is partially smoothed in the
value of the total radiative width of the cascade intermediate level by items
of different sign. On the other hand, experimental data allow one to account
to the fist approach for considerable enhancement in $k(E1)+k(M1)$ for the
secondary cascade transitions to the levels lying below the break threshold
of the second Cooper pair of nucleons.

This means that the conclusion \cite{0704-0916} on justice of hypothesis
\cite{BM} is not grounded. I. e., existing experimental data do not permit
one to exclude possibility of strong correlation between partial cross-sections
of gamma-transitions and evaporated nucleons for given excitation energies
of final nuclei.

2. Analysis of published results on study of two-step reaction
$^{95}$Mo$(n,2\gamma)$ demonstrates another aspect which should be taken into
account in order to obtain reliable conclusion about the studied picture of
processes occurring in nucleus.

So, impossibility to reproduce experimental intensity of two-step cascades
in limits of their total experimental error with the use of any model notions
of $\rho$ and $k$ or data of other experiments is absolute argument for their
more or less mistakenness. This is true within uncertainty of other
existing notions of gamma-decay process.
For instance, there is idea of independence of decaying
mode of excited levels on way of their excitation.

But, correspondence between calculation and experiment (moreover, in limited
interval of gamma-transition energy) cannot be a proof for justice of the
tested $\rho$ and $k$. This is due to their potential coincidence with one of
false solution from one hand and owing to impossibility to guess experimental
values of $\rho$ and $k$ - from other hand.

\section{Principles of two-step cascade analysis in $^{96}$Mo}\hspace*{16pt}
The goal of analysis suggested here is determination of the most reliable
$\rho$ and $k$ values for given isotope (independently on other opinions
concerning this point).

Unique (and not realized in \cite{TSC-96,PSF07-24}) possibility for this aim
 is provided by experimental data on two-step cascades measured in \v{R}e\v{z}.
  But, obtaining of reliable data is impossible without formulation of
   conditions and postulates of analysis providing its maximal reliability:

1. Cascade intensity is described by the function whose arguments have
infinite number of possible values. However, all these values are physically
limited by final interval of possible magnitudes for any energy of excitation
and gamma-transition:

\begin{eqnarray}
\rho_{min} <\rho <\rho_{max}\nonumber \\
\Gamma_{min} < \Gamma <\Gamma_{max}.
\end{eqnarray}

Therefore, the region of their possible values can be determined by
mathematical methods without using of model notions about $\rho$ and /or $k$.
But, this can be done only under condition that the ratio between strength
functions of primary and secondary gamma-transitions of the same energy
and multipolarity is set on basis of some hypotheses or experimental data.
This statement is easily tested using any algorithm of search for random
solution of system of equations (4) even at its spreading onto functional
dependence for cascades with reverse ordering of primary and secondary
gamma-transitions of considered energy. Of course, it is necessary
to set maximally different initial values for iterative process and,
it is desirable, out of the region (5) of its determined maximal and minimal
values. 

2. On the ground of experiments performed earlier, it must be assumed that
the cascade intensity depends on the wave function structures of its three
levels. If this statement is false for given nucleus then objective analysis
must demonstrate their independence.

3. The analysis must use all inherent to experiment possibilities.
Therefore, it is necessary to use the only found up to now possibility
\cite{PEPAN-2005} to estimate degree of functional dependence of strength
functions on energy (i.e., structure of wave function) of decaying level.

If nucleus $^{96}$Mo is exclusion from this rule, and the Axel-Brink hypothesis
is applicable and for it then experiment must show in limit of errors the
independence of wave functions on nuclear structure:
$k(E_\gamma , E_{ex})= k(E_\gamma)$.

\section{Results of analysis of two-step cascade intensities in
$^{96}$Mo}\hspace*{16pt}

As earlier, authors of \cite{TSC-96,PSF07-24} used for proof of their point
of view comparison between central parts of some experimental spectra
(corrected by efficiency of cascade registration) for choice of some variant
of model values of $\rho$ and $k$. In their opinion, the tested variants
describe the compound-state gamma-decay adequately to the experiment.

Practical use of the analysis principles enumerated above does not correspond
to this conclusion completely.


\begin{figure}[htbp]
\vspace{3cm}
\begin{center}
\leavevmode
\epsfxsize=12cm
\epsfbox{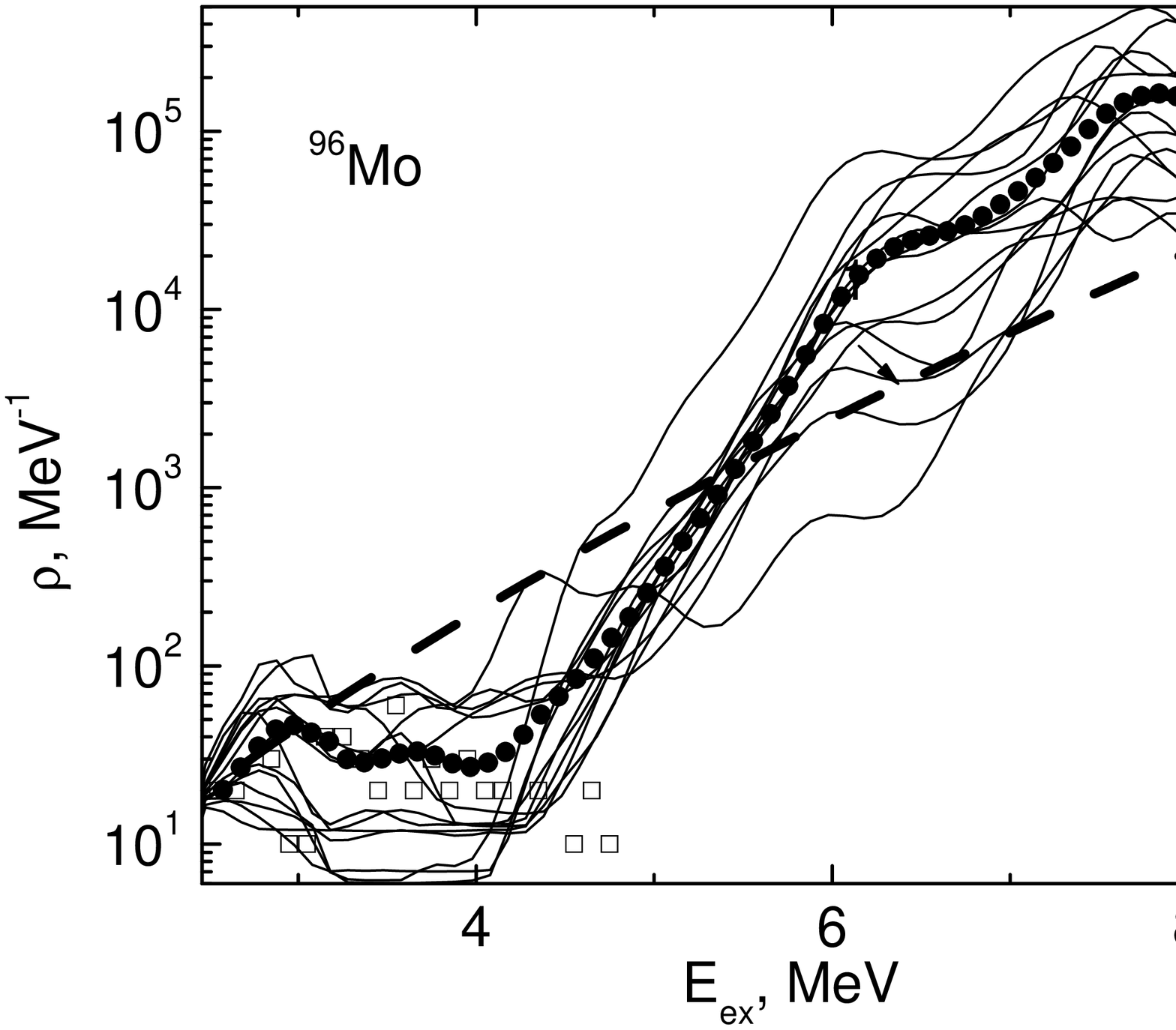}
\end{center}
\vspace{-4.5cm}
{ Fig.~3.} Curve  1 -- model values \cite{BSFG}, thin curves represent
the best random functions of the density of intermediate cascade levels
reproducing 
$I_{\gamma\gamma}$ in Fig. 5, 6 with practically the same values $\chi^2/f < 1$.
Solid points show their mean value.
Squares present data from \cite{ENSDF}.
\end{figure}

\begin{figure}[htbp]
\vspace{3cm}
\begin{center}
\leavevmode
\epsfxsize=12cm
\epsfbox{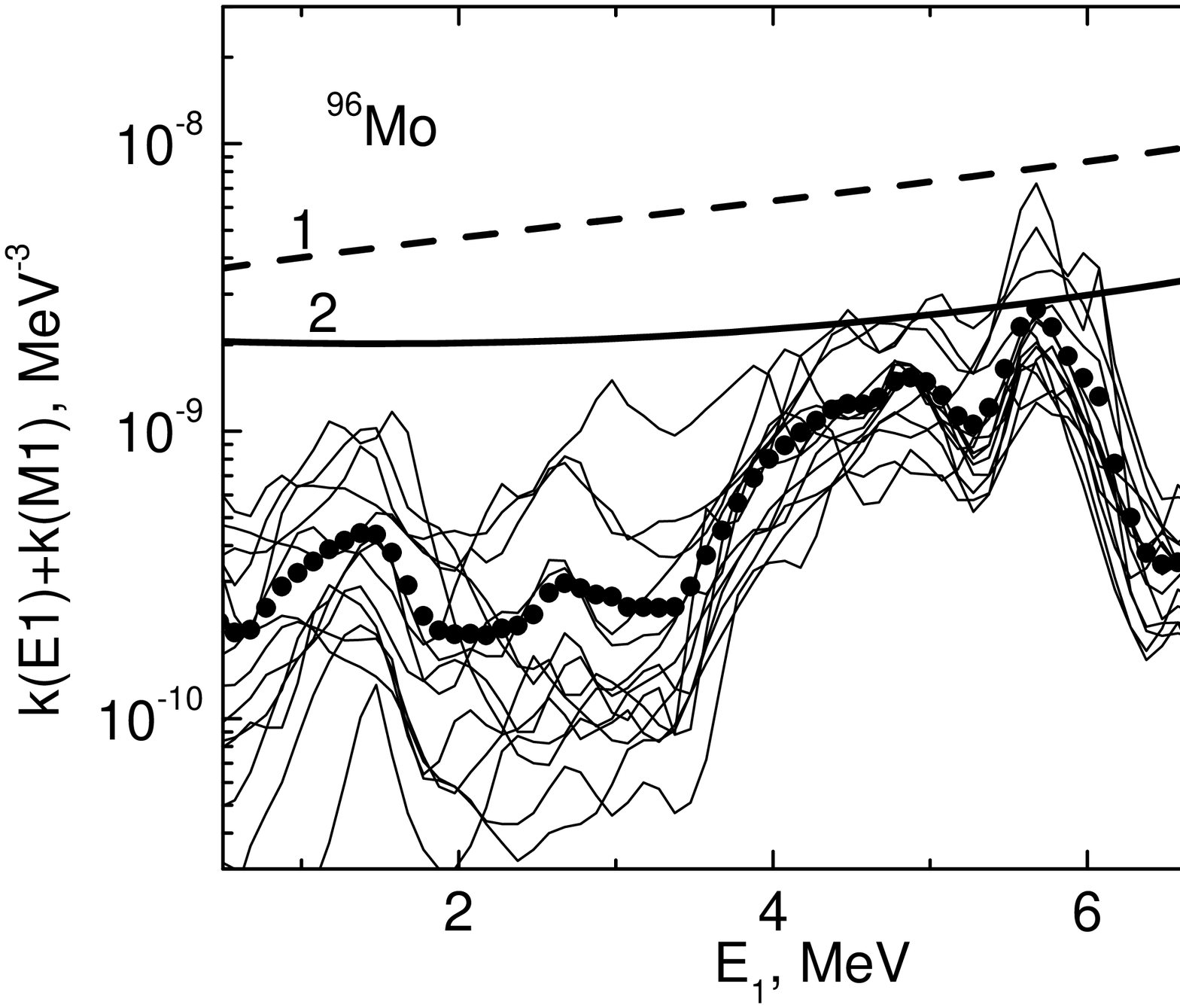}
\end{center}
\vspace{-4.5cm}
{Fig.~4.} Curve 1  - $k(E1)$ from model \cite{Axel}, curve 2 - \cite{KMF}
in sum with $k(M1)=$const. Thin curves represent the best random functions
reproducing  $I_{\gamma\gamma}$ in Fig. 5, 6 with practically the same values
$\chi^2/f <1$. Solid points show their mean value.
\end{figure}

In analysis of the experimental data performed by us were used the following
nuclear parameters:
density of levels with $J^{\pi}=2,3^{+}$ at $B_n$=9.154 MeV corresponds to
spacing between them $D$=55 eV. Below 2.5 MeV were used experimental scheme
of levels and modes of their decay.
Level density excited according to model \cite{BSFG} by primary dipole
transitions is presented in Fig. 3 by curve 1. It corresponds to nuclear
parameters enumerated above. Analysis in frameworks of this model gives level
density at the lowest and highest nuclear excitation energy.
It is fixed equality of level density with different parity at $B_n$.
Their ratio below neutron binding energy is free parameter.

The total radiative width of neutron resonances was taken equal to
$\Gamma_\gamma = 160$ meV, ratio $k(M1)/k(E1)$=0.256 for $E_\gamma=6.8$ MeV.
The threshold of fitted spectra was taken equal to 0.9 MeV for correspondence
with \cite{Morea}. The ratio of capture number in compound state with
$J=2^{+}$ to the total number of captures was accepted equal to $\sigma_J$=50
and 66\%. This variation was performed for possible compensation of error
in determination of spins (and capture cross-sections) in under-threshold
 resonances \cite{BNL-325}.

The functions presented in Fig. 1b) and 1c) were used as initial values of $\rho$ and $k$
in some part of calculation. And obviously unreal their values were used in
other part of calculation.


As it was already mentioned in \cite{TSC-err,Fe57,Au198}, the use of
experimental spectra with indefinite ratio between intensities of the primary
and secondary transitions in vicinity of the primary gamma-transition energy
$E_\gamma =E_1$ or $E_\gamma =E_2$ increases region of the possible $\rho$
and $k$ values by 1-2 order as compared with the used by us \cite{PEPAN-2005}
analysis. This is observed in figs. 3 and 4 (there is given for
$\sigma_J=66\%$).
Some decrease in dispersion of the found $\rho$ and $k$ values can be achieved
by the use of experimental data on primary transition intensities of the most
strong cascades. This is effectively in near-magic nuclei and, in principle,
allows one to reject clearly false ``bump" which appears itself in Fig. 4 about
$E_1=1.5$ MeV in some variants of calculation. This problem is solved
automatically in experimental data processing within the method
\cite{PEPAN-2005} if only the quanta ordering in the most intense cascades is
determined by means of apparatus of nuclear spectroscopy with errors not
exceeding some percents. However, possibility of mistaken determination of
quanta ordering in cascades with their intermediate level energies below
3-4 MeV in the considered isotope with relatively low level density is
practically unreal. Principle discrepancy in results of approximation with
variants $\sigma_J=50\%$ and $\sigma_J=66\%$ in spectra of possible $\rho$
and $k$ functions is not observed. Pure E2-transitions were not considered in
approximation; actually, the data in Fig. 4 may contain mixture of dipole and
quadrupole transitions.
\begin{figure}[htbp]

\vspace{3cm}
\begin{center}
\leavevmode
\epsfxsize=12cm
\epsfbox{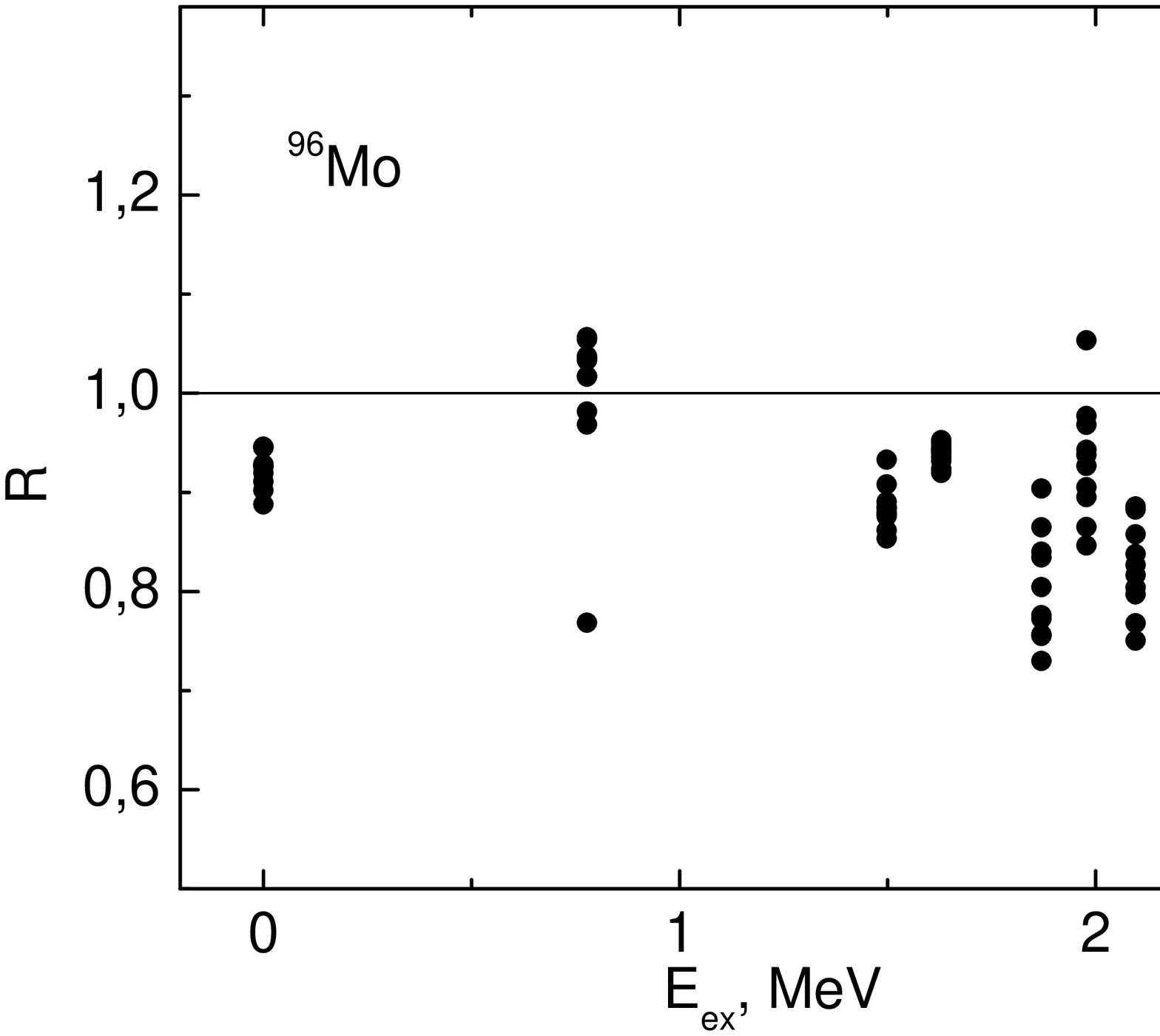}
\end{center}
\vspace{-4cm}

{\bf Fig.~5.}
The ratio between calculated and experimental intensities for central parts
of 11 spectra for random functions presented in figs. 3 and 4.
\end{figure}

The cascade intensity spectrum is given in \cite{PSF07-24} only for
four summed energies. Therefore, just these data were used in our fitting of
spectra. Results of approximation of summed intensity of central parts for
11 spectra  by random functions (figures 3 and 4) are presented in Fig. 5.
Resulting dispersion in calculated data presented in this figure for maximal
excitation energies can be stipulated by both errors in normalization of
spectra and errors in the used in calculation values of their spin and parity.

Anti-correlation between $\rho$ and $k$ results in fact that the level density
functions with their maximum values correspond to strength functions with the
least values (it directly follows from condition $\Gamma_\gamma =$ const).

\section{Conclusion}\hspace*{16pt}

1. Description of the experimental intensity of two-step cascades to a
precision of experiment is impossible in frameworks of many existing model
notions and experimental data.

2. Approximation of cascade intensity to the ground (phonon-less), one-phonon
($E_{ex}=778$) states and two-phonon doublet ($E_{ex}=1625+1629$) keV reflects,
probably, influence of wave function structure of cascade final level on its
intensity. Local ``bump" in region of the primary transition energy $\approx 4$
MeV determines the shape of their energy dependence and summed intensity of
each cascade.

3. More precise data on $\rho$ and $k$ can be obtained only by application of
methods \cite{Prim,PEPAN-2005} to the experimental data from Rez. Uncertainty
in determination of dependence (4) of cascade intensities on on energy
of their primary transitions within the method \cite{Prim} at given statistics,
resolution and background will be considerably less than error in normalization
of absolute intensity $I_{\gamma\gamma}$.

4. Spectroscopic data can permit one to determine cascade population of levels
in $^{96}$Mo up to the excitation energy not less than 4.5 MeV.
It is not excluded that this is quite enough for observation of increase
in strength functions of secondary transitions to the levels lying in region
$E_{ex} \approx 4$ MeV.

5. The shapes of energy dependences for $\rho$ and $k$ repeat, in the average,
analogous data for other nuclei -- step-like structure, strengthening of $k$
to the levels in its region not only for primary but, probably, also for
secondary cascade transitions, and noticeable decrease in low-energy region
of primary transitions $E_1 <3.5$ MeV.

6. The method used in Oslo needs in realistic estimation of systematical errors
of the measured spectra and their coefficient transfer to the determined
$\rho$ and $k$ values. It is not excluded that the required precision of
experiment estimated here cannot be achieved in existing variant of the
experiment even in principle.

7. Most probably, its authors would be able to obtain reliable enough data
by the use of two-step reaction ``charged particle + gamma-quantum to low-lying
level".

\begin{figure}[htbp]

\vspace{5cm}
\begin{center}
\leavevmode
\epsfxsize=17cm
\epsfbox{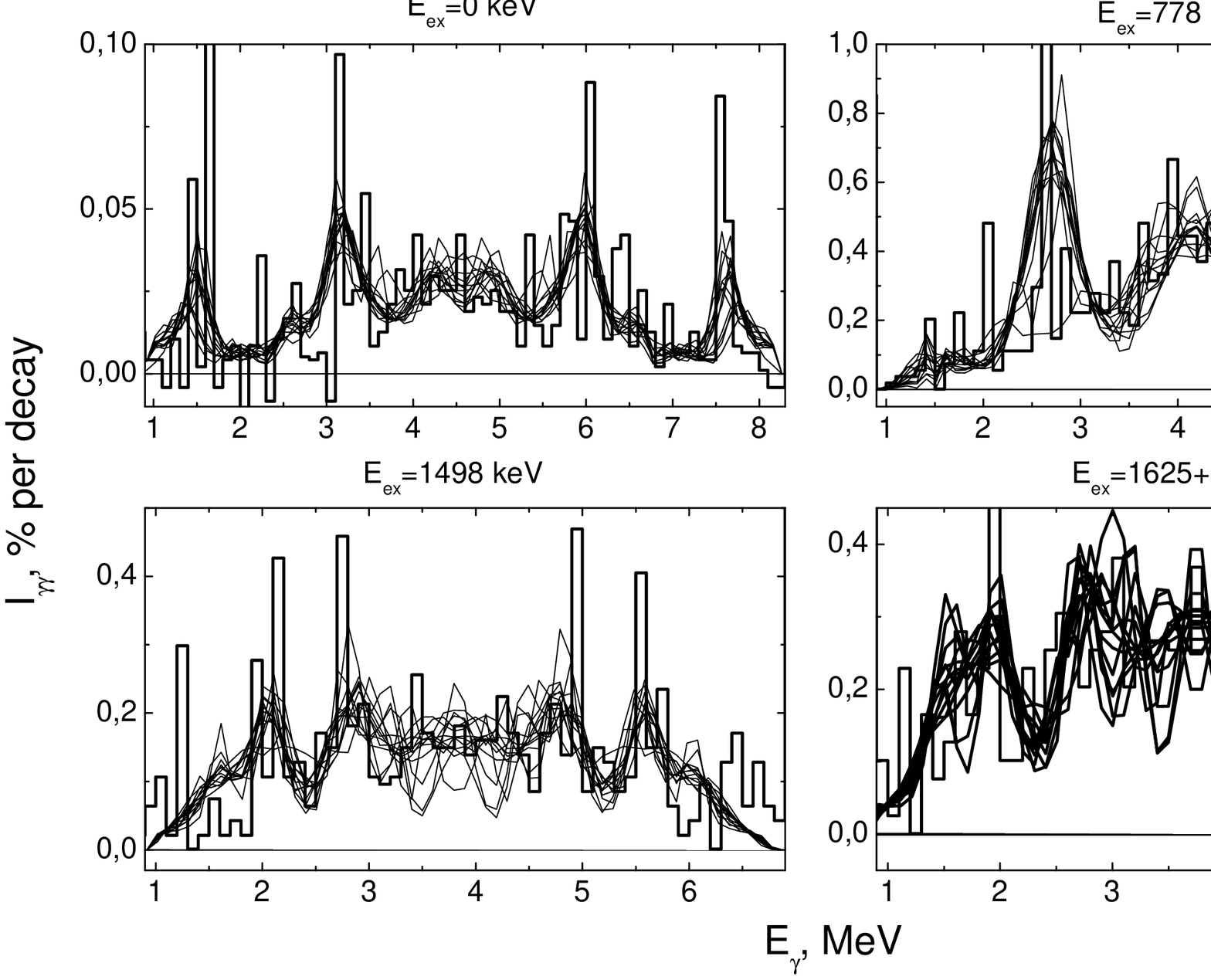}

\end{center}

\vspace{-5cm}
{\bf Fig.~6.} Histogram -- experimental intensity of two-step cascades for
the levels $E_{ex}$ (summed over the intervals of 100 keV).
Lines --  variants of the calculation with random level density and
radiative strength functions presented in figures 3 and 4.
Normalization of experimental and calculated spectra corresponds to that
adopted in \cite{PSF07-24}, i.e., corresponds to summed intensity of all
possible two-step cascades 200\% per decay.

\end{figure}

\end{document}